# The Dissipation of the Solar Nebula Constrained by Impacts and Core Cooling in Planetesimals

Alison C. Hunt[1], Karen J. Theis[2], Mark Rehkämper[3], Gretchen K. Benedix[4,5], Rasmus Andreasen[6] and Maria Schönbächler[1]

[1]Institute of Geochemistry and Petrology, ETH Zürich, Clausiusstrasse 25, 8092 Zürich, Switzerland.

[2]School of Earth and Environmental Sciences, The University of Manchester, Manchester, M13 9PL, UK.

Now at: The Photon Science Institute, Alan Turing Building, The University of Manchester, Manchester, M13 9PL, UK.

[33] Department of Earth Science & Engineering, Imperial College London, London, SW7 2AZ, UK.

[4]Space Science and Technology Centre, School of Earth and Planetary Sciences, Curtin University, GPO Box U1987, Perth, WA 6845, Australia.

[5]Department of Earth and Planetary Sciences, Western Australian Museum, Locked Bag 49 Welshpool DC, Western Australia 6986, Australia.

[6]Department of Geoscience, Aarhus University, Høegh-Guldbergs Gade 2, 8000 Aarhus C, Denmark.

*Correspondence and requests for materials should be addressed to: alison.hunt@erdw.ethz.ch; +41 (0)44 632 4643; +41 (0)44 632 1376 (fax)

**Abstract**

**Rapid cooling of planetesimal cores has been inferred for several iron meteorite parent bodies based on metallographic cooling rates, and linked to the loss of their insulating mantles during impacts. However, the timing of these disruptive events is poorly constrained. Here, we used the short-lived $^{107}$Pd-$^{107}$Ag decay system to date rapid core cooling by determining Pd-Ag ages for iron meteorites. We show closure times for the iron meteorites equate to cooling in the timeframe ~7.8-11.7 Myr after CAI, and indicate that an energetic inner Solar System persisted at this time. This likely results from the dissipation of gas in the protoplanetary disk, after which the damping effect of gas drag ceases. An early giant planet instability between 5–14 Myr after CAI could have reinforced this effect. This correlates well with the timing of impacts recorded by the Pd-Ag system for iron meteorites.**

**Main Text**

Iron meteorites are thought to represent the once heated interior portions of planetesimals and are some of the earliest-formed bodies in our Solar System[1]. As such, they are survivors of the many dynamical processes that have shaped Solar System architecture, including dissipation of the protoplanetary disk, and runaway growth, migration and reorganisation of the giant planets[2,3]. Importantly, high and variable metallographic cooling rates among many iron meteorite groups, including the IIAB, IIIAB and IVAs, indicate that these metals cooled in bodies that had been stripped of their insulating silicate mantles[4,5]. Dating core crystallisation can therefore elucidate the timing of impact processes, which in turn constrains the evolution of our early Solar System. The short-lived $^{107}$Pd-$^{107}$Ag decay system ($t_{1/2}$ ~6.5 Myr) provides a tool for this. Palladium, which is highly siderophile, partitions into the metal phase during crystallisation, and chalcophile Ag is strongly sequestered into sulfides. This leads to Pd-Ag fractionation during crystallisation of a cooling core resulting in samples from a single body with variable Pd/Ag ratios, thereby allowing us to derive parent-body initial

$^{107}$Pd/$^{108}$Pd ratios from isochron slopes. The ages for parent-body cooling through the closure temperature of the Pd-Ag system (~820-970 K)[6] can be derived from these initials.

Early studies of iron meteorites using the Pd-Ag chronometer did not fully account for the exposure of samples to galactic cosmic rays (GCR)[7-9], which can disturb the system through secondary neutron capture reactions by Pd isotopes that lead to a decrease in $^{107}$Ag/$^{109}$Ag. This occurs primarily via neutron capture on $^{108}$Pd and subsequent β⁻decay to $^{109}$Ag, and requires correction to achieve meaningful ages[10]. After GCR correction, initial $^{107}$Pd/$^{108}$Pd ratios of individual meteorites from the iron meteorite groups IIAB, IIIAB and IVA imply core cooling between ~7.2 Myr and 10.7 Myr after the formation of calcium-aluminium rich inclusions (CAIs)[6,11,12], relative to the Solar System initial (SSI) $^{107}$Pd/$^{108}$Pd determined from carbonaceous chondrites[13]. Additionally, a detailed investigation of the partially-differentiated IAB irons yielded two isochrons that signified a later closure of the Pd-Ag system between ~14.9–18.7 Myr after CAI[9]. However, those data were not corrected for GCR effects. Here, we focus on parent body initial $^{107}$Pd/$^{108}$Pd ratios to reassess the thermal evolution of planetesimals and constrain dynamic processes occurring in the early Solar System. We report new Pd-Ag data and use Pt isotopes to correct for the effects of GCR in the IAB, IIAB and IIIAB iron meteorites.

**Results and Pd-Ag ages**

We determined Pd-Ag and Pt isotope compositions for 10 IIAB and IIIAB iron meteorites and Pt isotope data for a further 8 IAB samples where Pd-Ag data were previously published[9] (Table 1; Methods; Supplementary Materials). Data are presented in the epsilon notation (i.e., $\varepsilon^{19i}$Pt/$^{195}$Pt is the deviation in parts per 10,000 from the average of two bracketing standards). Platinum isotope values for the iron meteorites range from ~0.0 to 0.4 for $\varepsilon^{196}$Pt (Table 1; Extended Data Fig. 1; Supplementary Materials) and compare well to published values[14-16]. Platinum isotope ratios fit well to the modelled effects for neutron capture, reaffirming that Pt isotopes are a powerful neutron dosimeter.

Corrections for GCR effects were applied to $\varepsilon^{107}$Ag (Methods; Table 2) and parent-body isochrons were calculated (Table 3; Extended Data Fig. 2). Corrected data define a single IAB isochron with a slope of 1.51 (±0.16) x10$^{-5}$, which is significantly different to the two previously determined isochrons[9]. The IIABs yield an isochron with a slope of 1.70 (±0.48) x10$^{-5}$, which is within uncertainty of previous GCR-corrected sample-specific isochrons for the group[11]. The IIIAB samples together define an isochron with a slope of 2.21 (±0.57) x10$^{-5}$, which overlaps with sample-specific $^{107}$Pd/$^{108}$Pd ratios determined for this group[6,11]. The GCR-corrected initial $^{107}$Pd/$^{108}$Pd ratios for the IABs, IIAB and IIIABs are within uncertainty of each other, indicating that the Pd-Ag system closed in a similar time-frame for these bodies (Fig. 1).

To determine Pd-Ag ages, the SSI $^{107}$Pd/$^{108}$Pd is needed. Two distinct initial $^{107}$Pd/$^{108}$Pd ratios have been determined based on the IVA iron meteorite Muonionalusta (~3.5 x10$^{-5}$; [12]) and carbonaceous chondrites (~5.9 x10$^{-5}$; [13]; see Methods for discussion of SSI). To consider the full range of SSI $^{107}$Pd/$^{108}$Pd ratios, we calculate Pd-Ag ages for the meteorite groups studied here relative to both (Table 3).

## Discussion

The presence of silicate inclusions and chemical data indicate the IAB iron meteorites had a complex thermal evolution and can be divided into sub-groups. The IABs studied here are members of the main group (MG) and 'Low-Au Low-Ni' (sLL) sub-group, which may be genetically related[17] (Table 1). It has been argued that the IABs formed in multiple impact melt pools on a single parent body[17]. Alternatively, they are products of internal heating followed by a catastrophic impact and reassembly of the parent body[18]. In support of the latter, samples from IAB sub-groups yield identical $^{182}$Hf-$^{182}$W ages indicating contemporaneous metal–silicate separation[14,19]. After the new correction for GCR effects, Pd-Ag data for the IABs define a single isochron indicating that cooling on the parent asteroid occurred in several sub-groups simultaneously. Small pools of metal formed at different times would not be expected to cool simultaneously and hence our new isochron likely represents a single body-wide cooling event, lending further support to a model of internal heating followed by breakup of the parent body.

Importantly, heating of the IAB parent asteroid produced a body which only partially differentiated at 6.0±0.8 Myr[14]. The question remains whether the body was impacted while at or close to its peak temperature, thereby effectively halting differentiation[18], or whether late accretion of the parent asteroid limited the availability of $^{26}$Al and allowed only subdued internal heating[14]. Thermal models explored a scenario where the body cooled slowly until ~10-12 Myr when the catastrophic breakup occurred[14]. However, the model used Pd-Ag data not corrected for GCR effects[9] to constrain the timing of the break-up, and hence a reassessment is necessary.

Our new IAB isochron corresponds to an age of 12.8 $^{+3.1}/_{-4.6}$ or 7.9 $^{+1.0}/_{-1.1}$ Myr after CAI (Table 3). Fast cooling rates are expected for the IAB parent body to keep silicate inclusions suspended in the metal portions[17]. Hence, a catastrophic impact should be closely followed by closure of the Pd-Ag system. Integrating this with cooling at 12.8 $^{+3.1}/_{-4.6}$ Myr suggests a delay between the body reaching its peak temperature at ~6 Myr and the catastrophic impact (Extended Data Fig. 3). Closure ages from the $^{129}$I-$^{129}$Xe system for a silicate inclusion from Campo del Cielo indicate a scrambling event that occurred later than ~11 Myr after CAI[20]. This agrees well with evidence of resetting from the Hf-W system at 14.9±2.8 and 11.4±2.2 Myr, respectively, in both the genetically related winonaite meteorites and a further silicate inclusion from Campo del Cielo[21-23]. This resetting must be impact-related and its identification in both winonaites and IABs indicates a large-scale process that likely corresponds to the catastrophic break-up of the body. Together with the Pd-Ag system these chronometers suggest an impact event between ~11 and 13.6 Myr after CAI. Assuming that the IAB parent was impacted at this time yields the necessary fast cooling rates of between instantaneous and ~10$^2$ K/Myr (Table 3; Methods). In this scenario, the thermal model and precursor parameters determined previously remain applicable to the IAB parent body[14].

Alternatively, a cooling age of 7.9 $^{+1.0}/_{-1.1}$ Myr (calculated relative to the SSI $^{107}$Pd/$^{108}$Pd of ~3.5 x10$^{-5}$) implies that the parent body was impacted when at or near its peak temperature (Extended Data Fig. 3). However, this scenario does not agree with the later resetting ages of silicate inclusions discussed above[20-22] because the closure temperature of the Hf-W system

(~1150 K[24]) is higher than that of the Pd-Ag system (~820-970 K[6]). Hence, a thermal event that affected the Hf-W chronometer should also reset the Pd-Ag system. This discrepancy is unlikely to represent disturbance to the Hf-W chronometer. Rapid cooling precludes significant diffusive exchange between silicate inclusions and host IAB metals[22]. Although GCR can affect Hf-W ages, this is unlikely to be the case here because Campo del Cielo irons experienced only limited GCR exposure (this study, [25]). A model incorporating cooling at 12.8 $^{+3.1}/_{-4.6}$ Myr is therefore most appropriate for the IAB parent body evolution.

**Implications for the Solar System initial $^{107}Pd/^{108}Pd$**

Importantly, the observation that the Hf-W and Pd-Ag systems cannot be reconciled using a SSI $^{107}Pd/^{108}Pd$ of ~3.5 x10$^{-5}$ is robust regardless of the formation history of the IAB parent body. Campo del Cielo is a member of the IAB-MG, which crystallised, and hence cooled, as a single body[17]. An isochron using only the IAB-MG yields an initial $^{107}Pd/^{108}Pd$ of (1.61±0.23) x10$^{-5}$, identical within uncertainty to the IAB isochron with all samples. Using a SSI $^{107}Pd/^{108}Pd$ of (3.5±0.1) x10$^{-5}$ implies closure of the Pd-Ag system for the MG at 7.3 $^{+1.3}/_{-1.5}$ Myr after CAI. Irrespective of whether this defines metal cooling after a local impact or a body-wide catastrophic event, the irons should then not include silicates with a Hf-W closure age of 11.4±2.2 Myr[22,23]. Therefore, we suggest that the SSI $^{107}Pd/^{108}Pd$ ratio of ~3.5 x10$^{-5}$ is too low and we focus mainly on the SSI ratio derived from carbonaceous chondrites in the following discussion.

**Evolution of iron meteorite parent bodies**

The Pd-Ag ages obtained here of 11.7 $^{+3.6}/_{-5.9}$ Myr after CAI for the IIABs, and 9.2 $^{+3.5}/_{-5.7}$ Myr for the IIIABs, suggest these bodies cooled in the same time period (Table 3; Fig. 1). The IVAs also yield a similar cooling age (7.8 $^{+3.0}/_{-4.4}$, [12]). The timings of metal-silicate differentiation and core cooling of an undisrupted asteroid are primarily a function of size and time of accretion[26]. The comparable accretion and core crystallisation ages[1] and cooling rates (Table 3; Methods) of these bodies may therefore imply that they had similar sizes. However, high and variable metallographic cooling rates among the IIAB, IIIAB and IVA irons indicate that these metals cooled in bodies that had been stripped of their insulating silicate mantles in 'hit-and-run' style impacts[4,5]. Intriguingly, we find that the IIAB, IIIAB and IVA asteroids were impacted in broadly the same timeframe (Fig. 1), suggesting an energetic Solar System environment persisted at this time.

**Early Solar System dynamics**

Nucleosynthetic isotope variations indicate a dichotomy between non-carbonaceous (NC) and carbonaceous-type (CC) asteroids[27]. The IIAB, IIIAB and IVA asteroids are NC and likely accreted in the inner Solar System[1]. All three indicate impact-related cooling and this provides evidence for energetic events with the ability to affect multiple locations in the inner disk at ~7.8-11.7 Myr after CAI. Additional evidence for an energetic state in the NC reservoir[25] includes the IAB impact event identified above that occurred in a similar timeframe, albeit $5.0^{+1.0}_{-1.1}$ Myr later than on the IVA body (Table 3). Other early-formed asteroids, including the H-chondrite parent body, the ungrouped Nedagolla iron meteorite, and the IID and IVB irons which originated in the outer disk (CC) reservoir, may also show evidence of impact processes in this interval[5,11,28-30]. Below, we discuss mechanisms that can lead to an energetic inner disk environment between ~7.8-11.7 Myr after CAI.

The presence and subsequent dissipation of the gas disk exerts one of the strongest controls on early Solar System orbital dynamics. The presence of gas in a protoplanetary disk dampens the eccentricities of planetesimals and stabilises the system. As the gas disperses, the damping effect of gas drag ceases and energetic collisions capable of disrupting planetesimals are predicted[31-33] (Fig. 2). Imaging of young stellar clusters suggest the overall lifetime of circumstellar disks is ~6–8 Myr[34], and the gas in our protoplanetary disk likely survived no more than ~4 Myr after CAI[1]. Whilst this is earlier than the impact timeframe discussed here, models of planetesimal dynamics after gas dispersion show that bodies initially have dynamically stable orbits, with eccentricity and inclination building up over several Myr[31,33]. Therefore, impacts occurring between ~7.8 and 11.7 Myr may be attributable to the dissipation of the gas disk. The first ~10-20 Myr after gas dissipation are the most important for both the number and mean velocity of collisions[31]. Further, this mechanism generates a chaotic asteroid belt for several Myr and thereby also provides a viable explanation for later impacts on the IAB parent body. The disk-wide nature of this mechanism also provides an explanation for impacts in the CC reservoir.

Giant planets also act in multiple ways to influence the orbits of nearby planetesimals. Rapid runaway growth is predicted to perturb the orbits of nearby planetesimals to high eccentricities and gravitationally scatter them into the inner Solar System, causing widespread mixing with the potential for more frequent collisions[35,36]. Additionally, Jupiter and Saturn are suggested to have undergone rapid gas-driven in- and then outward migration (the 'Grand Tack' model), leading to excitation of the asteroid belt and high velocity impacts in the inner Solar System[37,38]. The formation times of chondrules and their distinct nucleosynthetic fingerprints suggest the NC and CC reservoirs remained isolated until ~3-5 Myr after CAI, constraining the earliest time of giant planet migration[1,37]. However, impact events between ~7.8-11.7 Myr are more than ~3 Myr later than the runaway growth and hypothesised migration phase of the giant planets, and therefore cannot be attributed to these events. Additionally, these mechanisms are not expected to lead to bombardment events that lasted longer than ~0.5-1 Myr[37,39,40], whereas the energetic state of the inner Solar System recorded by the Pd-Ag system persisted for at least $3.9\,^{+2.3}_{-3.1}$ Myr after the closure of the IVA core (Table 3). This longevity problem persists if using the lower SSI $^{107}Pd/^{108}Pd$ ratio of ~3.5 x10$^{-5}$, which yields impacts at ~2.9-6.8 Myr after CAI that broadly coincide with the migration of the giant planets. In this case, a second mechanism such as runaway growth must also be invoked. Further, these two events together do not provide a framework for understanding the later impact recorded by the IABs.

Additionally, a giant planet instability (the 'Nice' model) could provide the necessary energy to the asteroid belt and inner Solar System (Fig. 2, Scenario B). This model can explain many features of our Solar System[2] and was initially assumed to occur later in Solar System history thereby providing an explanation for the Late Heavy Bombardment (LHB)[41]. However, recent studies have questioned both the evidence for a LHB and the dynamic consequences of delaying the instability[42-44]. An early giant planet instability occurring ≤100 Myr after gas dissipation better matches the architecture of our Solar System[42,45-48]. Although the precise timing of the instability is difficult to determine, models which predict the small size of Mars advocate an instability ~5–14 Myr after CAI[47,48]. Impacts on the iron parent bodies between ~7.8-11.7 Myr agree well with such a scenario and may provide the first evidence from the meteorite record for the occurrence and timing of an early giant planet instability. Such long-lived and Solar System wide events also provide an explanation for later impacts including on the IAB parent body, and impacts across the disk, including both the NC and CC reservoirs. Additionally, the effect of gas dissipation may act as the trigger for the early giant planet instability[49]. Hence, these two mechanisms may have acted together to create an energetic Solar System between ~7.8-11.7 Myr after CAI.

**Methods**

Eighteen aliquots of thirteen iron meteorites from the IAB, IIAB and IIIAB groups were analysed for this study (Table 1). Samples were cut using a CBN blade and subsequently polished using SiC paper to remove remaining weathered areas and fusion crust. Any visible troilite inclusions were physically removed from the metals. Palladium-silver isotope data for the IAB irons were published previously in [9], and Pt isotope data for Coahuila are from [15]. Material for Pd-Ag and Pt isotope analyses was taken from adjacent sampling locations, except for Negrillos, Caddo County and Canyon Diablo. Platinum isotope data for these samples are taken from the literature and indicate that they are unexposed or only weakly exposed to the effects of GCR (Table 1, [14,16]).

**Platinum isotope measurements**

Platinum isotope ratios were determined according to the methods described by [50]. In brief, samples were placed in ethanol in an ultrasonic bath for 5 minutes and then leached for 5 minutes in cold 2 M HCl before digestion with a 2:1 mix of concentrated $HNO_3$ and HCl (~8 ml $g^{-1}$ of sample). Samples were refluxed and dried, and then taken up in concentrated HCl. At this stage, 0.3 g aliquots for Pt isotope analyses were taken and dried. The aliquots were refluxed again with a 2:1 mix of concentrated $HNO_3$ and HCl, and then re-dissolved in preparation for the anion exchange procedure. This consisted of two stages and utilised Biorad AG® 1-X8 resin (200 – 400 mesh, chloride form). The first stage separated Pt from major matrix constituents, while the second stage separated Pt from Ir[50]. The second stage was repeated until samples had $^{191}Ir/^{195}Pt$ ratios ≤ 0.13, which is below the tolerance limit for Ir isotope tailing onto Pt isotopes during mass spectrometry, as determined for the instrument housed at ETH Zürich[50]. Osmium, which generates isobaric interferences on Pt isotopes, was then volatilised using perchloric acid.

Platinum isotope ratios were measured at ETH Zürich using a Thermo Scientific Neptune *Plus* operated in low-resolution mode and fitted with a Cetac Aridus II desolvating nebuliser and standard H cones, following the procedure of [50]. Samples were diluted to yield an ion beam intensity of ~2 x10$^{-10}$ A for $^{194}$Pt, which equates to ~150 ng g$^{-1}$ Pt and required ~170 ng per analysis. Analyses were corrected for instrumental mass bias using the exponential law and were internally normalized to $^{198}$Pt/$^{195}$Pt ('8/5') = 0.2145 [16]. Sample measurements were bracketed by the NIST SRM 3140 Pt standard solution and data are presented in the epsilon notation (i.e., $\varepsilon^{19i}$Pt/$^{195}$Pt = deviation in parts per 10,000 from the average of the two bracketing standards). Reproducibility for Pt isotope analyses was assessed using the iron meteorite North Chile as an in-house standard. Repeat analyses of multiple aliquots of this material indicate a 2 S.D. uncertainty of 0.73 for $\varepsilon^{192}$Pt, 0.15 for $\varepsilon^{194}$Pt and 0.09 for $\varepsilon^{196}$Pt (n=19; [50]). Platinum blanks are in the range 0.3 – 1.0 ng g$^{-1}$ sample.

**Pd-Ag Measurements**

Sample masses for Pd-Ag analyses are given in Supplementary Materials Table 1. Samples for Pd-Ag analyses were leached in cold concentrated aqua regia for 5 minutes, followed by digestion in concentrated aqua regia (~8 ml g$^{-1}$) on a hotplate. They were then dried and refluxed in concentrated HCl (40 ml), before adding 20 – 40 ml H$_2$O to verify that the sample was fully in solution. Approximately 9% of the sample solution was taken at this stage to determine Pd and Ag concentrations by isotope dilution, following an established method[8]. Samples were then refluxed and dried again in HCl to ensure no traces of HNO$_3$ remained. The ion exchange technique for Ag purification followed a published protocol[51] and utilised a three-stage column procedure with anion and cation exchange resin. The chemical procedure removes elements that cause interferences on Ag isotopes, including isobaric (Pd, Cd and Ru) and molecular interferences (Cu, Zn, Ga, Ge, Sr, Y, Zr, Nb and Mo), such that all were below the tolerance limits for accurate Ag isotope analyses[9,51]. The yields of the chemical separation procedure were checked for each sample individually and they were always close to 100%.

Silver isotope analyses were performed by MC-ICP-MS on either a Nu Plasma instrument in the School of Earth, Atmospheric and Environmental Science at The University of Manchester or a VG Axiom in the Department of Terrestrial Magnetism at the Carnegie Institution of Washington (Table 1). The details are described in [9] and [51], respectively, and a short summary is given here. Both instruments were operated with a glass cyclonic spray chamber. A Pd NIST SRM 3138 standard solution was added to all samples and standards just prior to analysis to correct for instrumental mass bias[51], and was mixed to yield a signal intensity ratio of 1 for $^{107}$Ag/$^{108}$Pd. Additionally, sample measurements were bracketed by the NIST SRM 978a Ag - NIST SRM 3138 Pd standard solution mixture, with the Pd and Ag concentrations of the standard matched to the samples to within 20% or better. Instrumental mass bias was corrected using the exponential law with normalisation to a $^{108}$Pd/$^{105}$Pd value of 1.18899 [7]. Data are presented in the epsilon notation and a $^{107}$Ag/$^{109}$Ag ratio of 1.079760 [13] was used for the NIST SRM 978a Ag standard.

For analyses conducted at The University of Manchester, the typical daily reproducibility of a 100 ppb Ag standard solution for the $^{107}$Ag/$^{109}$Ag ratio was ~0.25 ε (2 S.D.[9]. Additionally, the external reproducibility of sample measurements was assessed by repeatedly passing the NIST SRM 978a Ag standard, the USGS standard Cody Shale SCo-1, and the Hawaiian basalt KOO49 through the chemical procedure. These materials yielded external reproducibilities of between 0.1 and 0.6 ε (2 S.D.[9]), and therefore a conservative external uncertainty of ±0.6 ε has been applied to all sample data acquired at The University of Manchester. Similar tests were carried out at the Carnegie Institution[51] and resulted in an external uncertainty of ±0.5 ε . Silver blanks for sample digestion and column chemistry were <0.254 ng.

**Correction for Galactic Cosmic Ray Effects to Ag isotopes**

New ε$^{107}$Ag data for the IIAB and IIIAB irons vary from ~37 to 103, while $^{108}$Pd/$^{109}$Ag ratios range between 261 and 679 (Table 1). There is a positive correlation between ε$^{107}$Ag and $^{108}$Pd/$^{109}$Ag for all meteorite groups, however, a regression cannot be fitted through our new data for the IIAB or IIIAB iron meteorites before corrections for GCR (Extended Data Fig. 2). The uncorrected IAB samples define an initial $^{107}$Pd/$^{108}$Pd ratio of 1.06 (± 0.36) x10$^{-5}$ with a MSWD of 45 [9].

Galactic cosmic rays (GCR) can disturb the $^{107}$Pd-$^{107}$Ag chronometer through reactions with thermal neutrons[10]. This must be corrected before meaningful age information can be obtained. Secondary neutron capture by Pd isotopes leads to a decrease in $^{107}$Ag/$^{109}$Ag primarily via neutron capture on $^{108}$Pd and subsequent β⁻ decay to $^{109}$Ag. Additionally, $^{106}$Pd is also susceptible to neutron capture followed by β⁻ decay to $^{107}$Ag. The GCR-induced shift in $^{107}$Ag/$^{109}$Ag is therefore linearly correlated with Pd/Ag and higher $^{108}$Pd/$^{109}$Ag ratios lead to larger negative shifts in $^{107}$Ag/$^{109}$Ag, i.e., the slope of data not corrected for GCR effects would be too shallow, if one could even be determined because of the scatter[10]. Longer exposure ages, bigger pre-atmospheric radii, and increasing shielding depths all lead to larger GCR offsets.

It has been demonstrated that Pt isotopes are a powerful neutron dosimeter for correcting the effects of GCR in short-lived decay systems, e.g., $^{182}$Hf-$^{182}$W [14]. They have also been applied as a dosimeter for neutron capture in the $^{107}$Pd-$^{107}$Ag system[11,12]. New Pt isotope data fit well to the modelled effects for neutron capture (Extended Data Fig. 1). Model trendlines are shown for Ir/Pt ratios appropriate for our samples (Supplementary Materials Table 1). These results further confirm that neutron capture reactions are the cause of Pt isotope excesses and that there are no detectable nucleosynthetic isotope variations between iron meteorites.

Following the method of [11] (given as equations 1-3 below), data for the IAB, IIAB and IIIAB irons were assessed for the effects of neutron capture. The correction relies on the $^{108}$Pd/$^{109}$Ag ratio and neutron dose ($\varepsilon^{196}$Pt) of the sample in question. Measured $\varepsilon^{196}$Pt can be related to neutron capture corrections to $\varepsilon^{107}$Ag ($\varepsilon^{107}$Ag$_{GCR}$) by a linear regression:

$$\varepsilon^{107}Ag_{GCR} = \textit{Offset} + \textit{Slope} \times \varepsilon^{196}Pt \quad\quad\quad\quad (1)$$

where *Offset* and *Slope* are functions of the measured $^{108}$Pd/$^{109}$Ag ratio[11]. The *Offset* defines neutron capture shifts on Ag isotopes and is described by:

$$\text{Offset} = -0.023 + 0.0024 \times (^{108}\text{Pd}/^{109}\text{Ag}) - 4.76 \times 10^{-7} \times (^{108}\text{Pd}/^{109}\text{Ag})^2 \quad (2)$$

while the *Slope* is defined by:

$$\text{Slope} = 1.539 - 0.226 \times {}^{108}\text{Pd}/^{109}\text{Ag} \quad (3)$$

The *Offset and Slope* are then used in Eq. (1) to determine $\varepsilon^{107}\text{Ag}_{\text{GCR}}$. For the samples included in this study $\varepsilon^{107}\text{Ag}_{\text{GCR}}$ ranges from 2.3 to -49.2, with the largest offsets recorded in the IIIAB irons (Table 2).

**Solar System Initial $^{107}\text{Pd}/^{108}\text{Pd}$**

In order to accurately determine Pd-Ag ages, the Solar System initial (SSI) $^{107}\text{Pd}/^{108}\text{Pd}$ is needed. Recent estimates are based on either carbonaceous chondrites[13] or the IVA iron Muonionalusta[11,12], which has resulted in a range of initial ratios. By equating the $^{53}\text{Mn}$-$^{53}\text{Cr}$ and $^{107}\text{Pd}$-$^{107}\text{Ag}$ decay systems, which both date volatile depletion, carbonaceous chondrites define a SSI $^{107}\text{Pd}/^{108}\text{Pd}$ of $(5.9\pm2.2) \times 10^{-5}$ [13]. This ratio is among the highest initials yet determined and includes a relatively large uncertainty. Values for the initial $^{107}\text{Pd}/^{108}\text{Pd}$ based on Muonionalusta range from $\sim(2.8–6.6) \times 10^{-5}$ [11,12], but the use of this meteorite to define a SSI $^{107}\text{Pd}/^{108}\text{Pd}$ is hampered by ambiguity over its absolute Pb-Pb age, determined from a troilite[52]. Recently, the Pd-Ag systematics of Muonionalusta were re-examined and two possible SSI $^{107}\text{Pd}/^{108}\text{Pd}$ ratios of $(3.5\pm0.1)$ and $(6.6\pm0.4) \times 10^{-5}$ were determined[12], dependent on the $^{238}\text{U}/^{235}\text{U}$ ratio of the Muonionalusta troilite. In order to consider the full range of SSI values, we calculate Pd-Ag ages for the iron meteorite groups studied here relative to both the lowest recent estimate based on Muonionalusta[12] ($\sim3.5 \times 10^{-5}$) and the value indicated from carbonaceous chondrites[13] ($\sim5.9 \times 10^{-5}$), which also encompasses the highest ratio estimated from Muonionalusta. To circumvent the ambiguity over the SSI $^{107}\text{Pd}/^{108}\text{Pd}$, closure times can also be calculated relative to the value determined for the IVA irons ($^{107}\text{Pd}/^{108}\text{Pd} \sim 2.57 \times 10^{-5}$, [12]; Table 3).

**Calculation of cooling rates**

The closure temperature of the Pd-Ag system is dependent on cooling rate, Ni and FeS content of the meteorite, and is estimated to be in the range ~820 - 970 K for the iron meteorites studied here[6]. Fast cooling rates are necessary for the IAB parent body in order to keep silicate inclusions suspended in the metal portions[17]. Assuming that the IAB parent was at its peak temperature of 1470 K[14] when impacted at ~11 Myr (see Main Text) yields the expected fast cooling rates of between instantaneous and ~102 K/Myr (Table 3). Alternatively, a cooling age of 7.9 $^{+1.0}/_{-1.1}$ (calculated relative to an initial SSI $^{107}Pd/^{108}Pd$ of ~3.5 x 10$^{-5}$ [12]) implies that the parent body was impacted when at or near its peak temperature at 6.0 ± 0.8 Myr. This yields cooling rates of between instantaneous and ~135 K/Myr.

Modelling indicates the IIAB and IIIAB parent bodies accreted < ~0.4 Myr after CAI[1]. Core formation on the IIAB parent body followed at 0.7 ± 0.3 Myr, based on the Hf-W chronometer[16]. Assuming a liquidus temperature of ~1570 K for the IIABs[53] and a steady rate of cooling between core formation and closure of the Pd-Ag chronometer indicates cooling rates of ~278-69 K/Myr for cooling at 6.8 $^{+2.3}/_{-3.1}$ Myr, or ~156-40 K/Myr for cooling at 11.7 $^{+3.6}/_{-5.9}$ Myr (Table 3). Metallographic cooling rates for the IIABs suggest a much slower process (0.8-10 K/Myr[54]). However, it is important to note that these apply to cooling at ~670-770 K, and that newer techniques have led to upward revisions of rates by factors of 10-50[5].

The IIIABs experienced core formation at 1.2 ± 0.3 Myr with a predicted liquidus temperature of ~1650 K[16,53]. This yields a very wide range of cooling rates, from instantaneous-121 K/Myr for Pd-Ag closure at 4.3 $^{+2.2}/_{-2.8}$ Myr to 415-58 K/Myr for closure at 9.2 $^{+3.5}/_{-5.7}$ Myr (Table 3). Cooling rates determined here for the IIIAB irons compare well with those derived previously using the Pd-Ag system[6,11]. They also overlap with recent estimates of metallographic cooling rates obtained for this body (338 – 56 K/Myr [5]), particularly for a Pd-Ag closure time of ~9.2 Myr after CAI.

**Acknowledgements**

This work was supported by the European Research Council under the European Union's Seventh Framework Programme (FP7/2007–2013/ERC Grant agreement n˚ [279779], MS). We gratefully acknowledge funding from STFC (ST/F002157/1 and ST/J001260/1, MR; ST/J001643/1, MS) and funding from the Swiss National Science Foundation (Project 200020_179129, MS). AH wishes to thank Mattias Ek and Manuela Fehr for laboratory assistance at ETH Zürich during this study. MS would like to thank Rick Carlson and Mary Horan (DTM, Carnegie Institution) for their support and the opportunity to analyse Ag isotopes in May 2006. We also thank Caroline Smith and Deborah Cassey (Natural History Museum, London), and Julie Hoskin (Smithsonian Institution National Museum of Natural History) for the loan of samples used in this work.

**Author Contributions**

M.S. designed the study. A.H., K.T. and M.S. prepared samples for isotope analyses and conducted the isotopic measurements. All authors were involved in the data interpretation and writing of the manuscript.

**Competing Interests Statement**

The authors declare no competing interests.

**Supplementary Information** is available for this paper in the form of one Table as a pdf.

**Data Availability**

All data are available within this manuscript and its Supplementary information, or available from the authors on request.

**TABLES, MAIN ARTICLE**

**Table 1.** Concentration data for Pd and Ag, isotope data for Ag, and $\varepsilon^{196}$Pt for IAB, IIAB and IIIAB iron meteorites.

| Meteorite | Class [a] | Pd (ppb) | Ag (ppb) | $^{108}$Pd/$^{109}$Ag | $\varepsilon^{107}$Ag | ±2 S.D. | $^{107}$Ag/$^{109}$Ag | $\varepsilon^{196}$Pt | ±2 S.D. |
|---|---|---|---|---|---|---|---|---|---|
| *Terrestrial* | | | | | | | 1.07976 [b] | | |
| Campo del Cielo [c] | IAB-MG | 3616 | 11.3 | 177.8 | 15.8 | 0.6 | 1.081466 | 0.09 | 0.09 |
| Canyon Diablo [c] | IAB-MG | 3888 | 39.0 | 53.9 | 1.4 | 0.6 | 1.079911 | 0.00 [e] | 0.09 |
| Odessa 1 [c] | IAB-MG | 4199 | 28.2 | 81.1 | 2.2 | 0.6 | 1.079998 | 0.31 | 0.09 |
| Odessa 2 [c] | IAB-MG | 4063 | 10.9 | 199.8 | 11.7 | 0.6 | 1.081023 | 0.28 | 0.09 |
| Goose Lake 1 [c] | IAB-sLL | 4472 | 22.4 | 109.9 | 7.4 | 0.6 | 1.080559 | 0.08 | 0.09 |
| Goose Lake 2 [c] | IAB-sLL | 3881 | 18.5 | 116.6 | 8.6 | 0.6 | 1.080689 | 0.08 | 0.09 |
| Toluca [c] | IAB-sLL | 4546 | 38.3 | 64.7 | 1.4 | 0.6 | 1.079911 | 0.13 | 0.09 |
| Caddo County [c] | IAB-UG | 4113 | 58.9 | 37.7 | 0.1 | 0.6 | 1.079771 | -0.01 [e] | 0.09 |
| Coahuila 1 | IIAB | 1829 | 3.3 | 307.9 | 53.4 | 0.6 | 1.085530 | 0.01 [f] | 0.07 |
| Coahuila 2 [d] | IIAB | 1708 | 3.8 | 251.2 | 50.6 | 0.5 | 1.085223 | 0.01 [f] | 0.07 |
| Negrillos [d] | IIAB | 1587 | 3.4 | 260.9 | 52.7 | 0.5 | 1.085450 | -0.03 [g] | 0.05 |
| North Chile 1 | IIAB | 1932 | 3.3 | 324.4 | 55.0 | 0.6 | 1.085697 | 0.12 | 0.08 |
| North Chile 2 [d] | IIAB | 1841 | 3.2 | 321.6 | 53.1 | 0.5 | 1.085488 | 0.12 | 0.08 |
| Sikhote Alin 1 | IIAB | 2451 | 3.2 | 433.5 | 59.9 | 0.6 | 1.086227 | 0.22 | 0.09 |
| Sikhote Alin 2 | IIAB | 2649 | 4.8 | 310.4 | 36.9 | 0.6 | 1.083746 | 0.22 | 0.09 |
| Boxhole | IIIAB | 2385 | 1.9 | 678.7 | 103.5 | 0.6 | 1.090933 | 0.25 | 0.09 |
| Henbury | IIIAB | 2199 | 2.0 | 588.8 | 77.7 | 0.6 | 1.088147 | 0.38 | 0.09 |
| Thunda | IIIAB | 3092 | 4.1 | 407.8 | 83.3 | 0.6 | 1.088754 | 0.05 | 0.09 |

Uncertainties for $\varepsilon^{107}$Ag are based on repeat analyses of reference materials (see Methods). Uncertainties (2 S. D.) on $^{108}$Pd/$^{109}$Ag ratios are estimated to be ~1 % for data collected at the University of Manchester and 5 % for data collected at the Carnegie Institution of Washington. Platinum isotope uncertainties are based on repeat analyses of an in-house standard (see Methods). [a] IAB sub-groups from [17]. See main text for definitions. [b] Terrestrial $^{107}$Ag/$^{109}$Ag ratio from [13]. [c] Palladium-silver data taken from [9]. All Pd-Ag isotope data measured at the University of Manchester, except [d] Carnegie Institution of Washington. Platinum isotope data taken from [e] Hunt, et al. [14], [f] Hunt, et al. [15] and [g] Kruijer, et al. [16].

**Table 2.** Correction to $\varepsilon^{107}$Ag for GCR effects for IAB, IIAB and IIIAB iron meteorites.

| Meteorite | $\varepsilon^{107}Ag_{meas}$ | ±2 S.D. | $\varepsilon^{107}Ag_{GCR}$ | ± 2 S.D. | $\varepsilon^{107}Ag_{pre-exp}$ | ± 2 S.D. |
|---|---|---|---|---|---|---|
| *IAB* | | | | | | |
| Campo del Cielo | 15.8 | 0.6 | -3.1 | 3.5 | 18.9 | 3.5 |
| Canyon Diablo | 1.4 | 0.6 | 0.1 | 1.0 | 1.3 | 1.1 |
| Odessa 1 | 2.2 | 0.6 | -5.0 | 1.5 | 7.2 | 1.6 |
| Odessa 2 | 11.7 | 0.6 | -11.9 | 3.9 | 23.6 | 4.0 |
| Goose Lake 1 | 7.4 | 0.6 | -1.6 | 2.1 | 9.0 | 2.2 |
| Goose Lake 2 | 8.6 | 0.6 | -1.7 | 2.2 | 10.3 | 2.3 |
| Toluca | 1.4 | 0.6 | -1.6 | 1.2 | 3.0 | 1.3 |
| Caddo County | 0.1 | 0.6 | 0.1 | 0.6 | 0.0 | 0.9 |
| | | | | | | |
| *IIAB* | | | | | | |
| Coahuila 1 | 53.4 | 0.6 | -0.2 | 4.8 | 53.7 | 4.8 |
| Coahuila 2 | 50.6 | 0.5 | -0.2 | 3.9 | 50.8 | 3.9 |
| Negrillos | 52.7 | 0.5 | 2.3 | 2.9 | 50.4 | 2.9 |
| North Chile 1 | 55.0 | 0.6 | -8.2 | 5.7 | 63.2 | 5.7 |
| North Chile 2 | 53.1 | 0.5 | -8.2 | 5.6 | 61.2 | 5.6 |
| Sikhote Alin 1 | 59.9 | 0.6 | -20.7 | 8.7 | 80.6 | 8.7 |
| Sikhote Alin 2 | 36.9 | 0.6 | -14.7 | 6.2 | 51.6 | 6.2 |
| | | | | | | |
| *IIIAB* | | | | | | |
| Boxhole | 103.5 | 0.6 | -37.1 | 13.7 | 140.6 | 13.7 |
| Henbury | 77.7 | 0.6 | -49.2 | 11.9 | 126.8 | 11.9 |
| Thunda | 83.3 | 0.6 | -3.7 | 8.2 | 87.0 | 8.2 |

**Table 3.** Parent body initial $^{107}Pd/^{108}Pd$ ratios and closure times of the Pd-Ag system relative to CAI and IVA irons.

| Group | Parent Body Initial $^{107}Pd/^{108}Pd$ (x $10^{-5}$) | Time after CAI (Myr) [a] | Time after CAI (Myr) [b] | Time after IVA (Myr) [c] | Cooling rates (K/Myr) [b] |
|---|---|---|---|---|---|
| IAB | 1.51 (± 0.16) | 12.8 $^{+3.1}$ / $_{-4.6}$ | 7.9 $^{+1.0}$ / $_{-1.0}$ | 5.0 $^{+1.0}$ / $_{-1.1}$ | Instantaneous – 102 |
| IIAB | 1.70 (± 0.48) | 11.7 $^{+3.6}$ / $_{-5.9}$ | 6.8 $^{+2.3}$ / $_{-3.1}$ | 3.9 $^{+2.3}$ / $_{-3.1}$ | ~156 – 40 |
| IIIAB | 2.21 (± 0.57) | 9.2 $^{+3.5}$ / $_{-5.7}$ | 4.3 $^{+2.2}$ / $_{-2.8}$ | 1.4 $^{+2.2}$ / $_{-2.8}$ | ~415 - 58 |
| IVA [c] | 2.57 (± 0.07) | 7.8 $^{+3.0}$ / $_{-4.4}$ | 2.9 $^{+0.4}$ / $_{-0.4}$ | - | > 500 [c] |

[a] Time after CAI and resultant cooling rates calculated using a SSI $^{107}Pd/^{108}Pd$ of 5.9 (± 2.2) x $10^{-5}$ [13]. [b] Time after CAI calculated using a SSI $^{107}Pd/^{108}Pd$ of 3.5 (± 0.1) x $10^{-5}$ [12]. See Extended Methods for calculation of cooling rates. [c] Initial $^{107}Pd/^{108}Pd$ and cooling rate for the IVA iron meteorites from [12].

**FIGURE LEGENDS, MAIN ARTICLE**

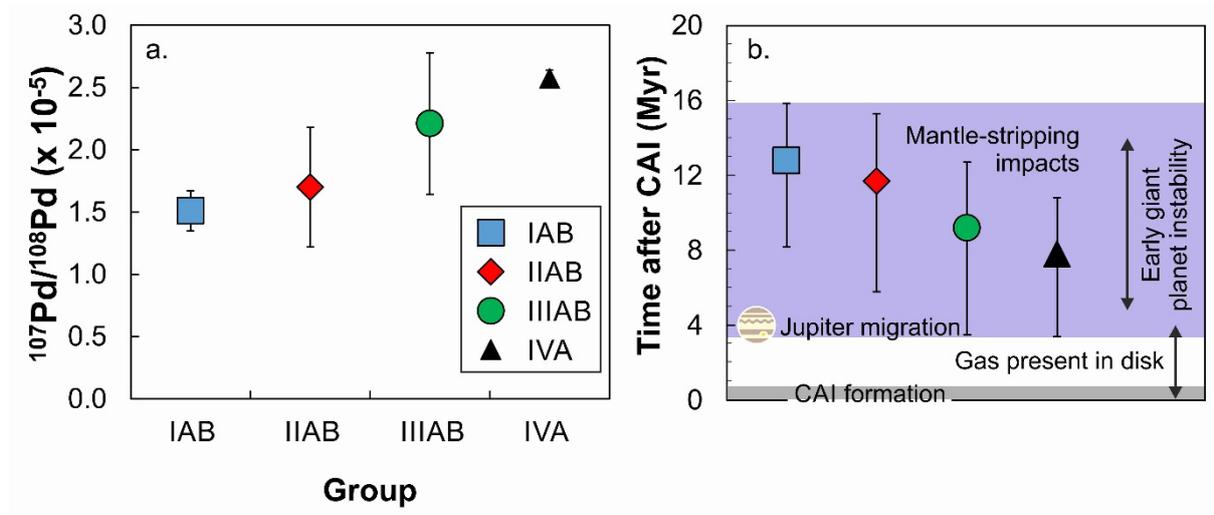

**Figure 1.** The timing of core cooling and impacts constrained by the Pd-Ag system. a. Parent body initial $^{107}Pd/^{108}Pd$ ratios for the IAB, IIAB and IIIAB iron meteorite groups (Table 3). Literature data for the IVA irons[12] are also shown. b. Age after CAI for core cooling, calculated relative to the carbonaceous chondrite SSI[13]. Iron meteorite group symbols as in panel (a). The grey and purple bands represent, respectively, the approximate timings for CAI formation and the impacts recorded by the Pd-Ag system. The timeframe suggested for the 'Grand Tack' is indicated by the size of the Jupiter symbol[1,37]. The intervals suggested for gas present in the disk, and an early giant planet instability that predicts a small Mars[47,48], are also shown.

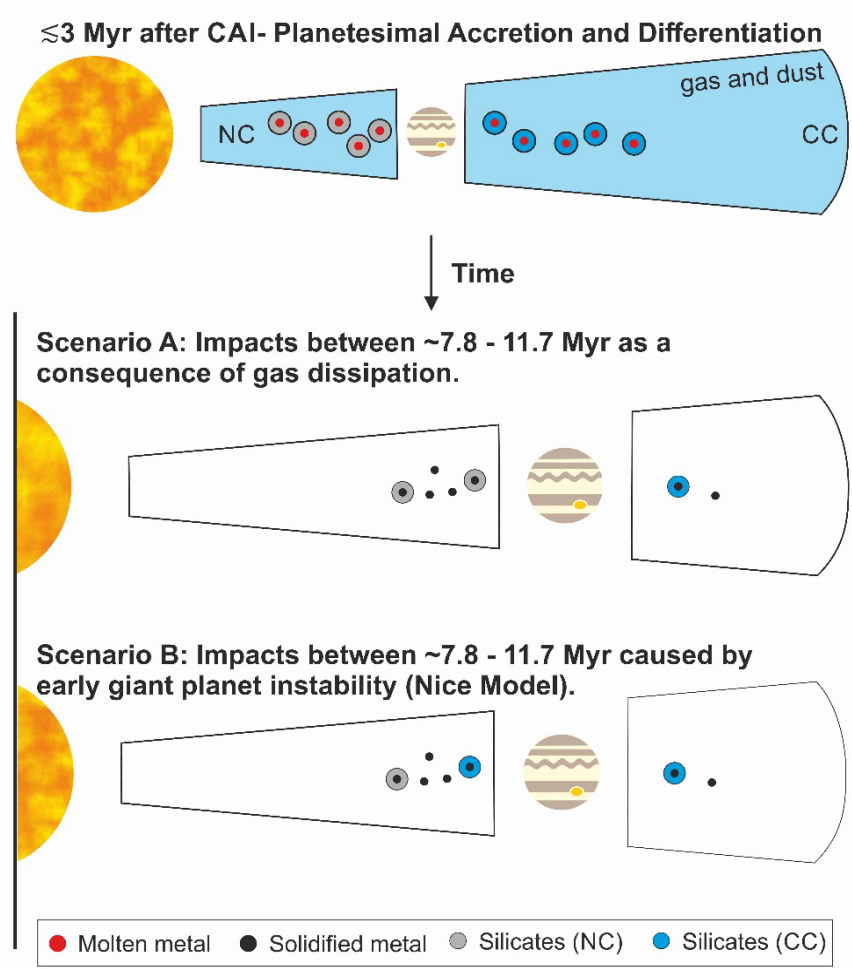

**Figure 2.** Cartoon illustrating the evolution of differentiated iron meteorite bodies in the early Solar System (not to scale). Top panel: Parent bodies accrete and differentiate within the first ~3 Myr after CAI[16]. The disk consists of gas and dust (coloured blue) and is truncated in the region of Jupiter, preventing significant exchange between the non-carbonaceous (NC) and carbonaceous (CC) reservoirs. Later, the cores of the IIAB, IIIAB and IVA bodies were exposed by impact events, leading to fast cooling and closure of the Pd-Ag system between ~7.8 - 11.7 Myr. Two scenarios, acting alone or in combination, provide mechanisms for an energetic inner Solar System. Scenario A: When the gas disk dissipates, planetesimals are gradually excited until they begin to undergo high velocity impacts[31]. Gas dissipation is completed before this time. Scenario B: An early giant planet instability is triggered at ~7.8 - 11.7 Myr, leading to reorganisation of the inner Solar System and impact events[47].

**EXTENDED DATA FIGURE LEGENDS**

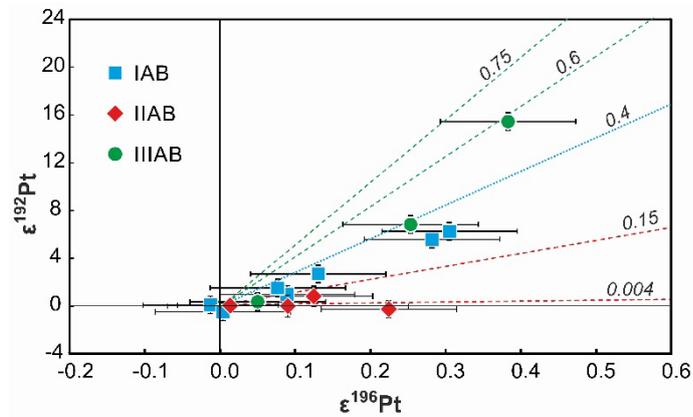

**Extended Data Figure 1.** Covariation of ε$^{192}$Pt and ε$^{196}$Pt for the IAB, IIAB and IIIAB iron meteorites. The GCR model calculations[10] are shown for Ir/Pt ratios measured for these groups (dashed lines; Ir/Pt ratios for individual samples are given in Supplementary Materials Table 1). Uncertainties are 2 S.D.

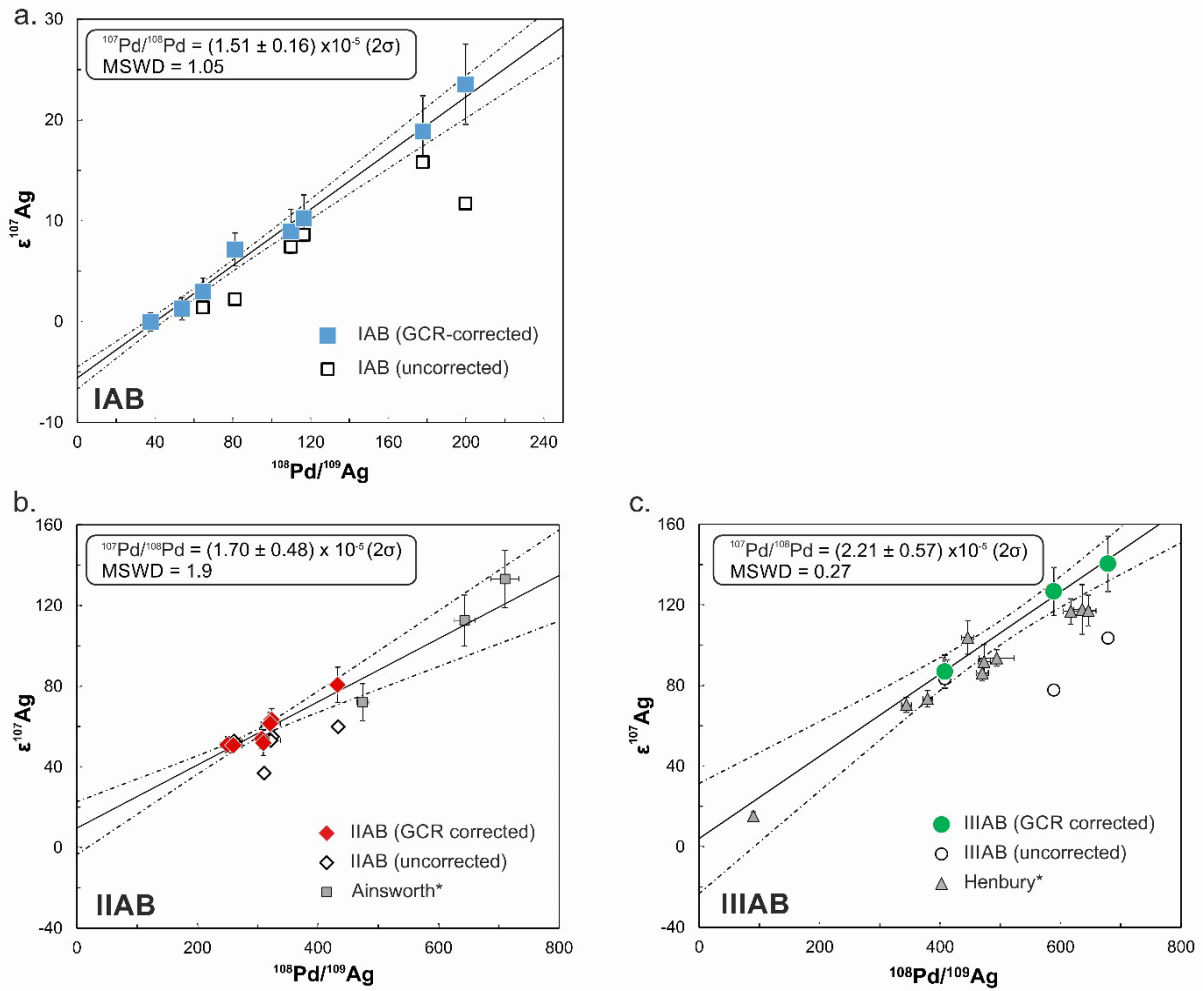

**Extended Data Figure 2.** Isochron diagrams for a) IAB, b) IIAB, and c) IIIAB iron meteorites for the Pd-Ag system. In each case, $\varepsilon^{107}$Ag is plotted against $^{108}$Pd/$^{109}$Ag for both GCR-corrected (filled symbols) and GCR-uncorrected data (unfilled symbols). Uncertainties on GCR-uncorrected data are 2 S.D. and within the size of the symbol (Table 1), while uncertainties for $\varepsilon^{107}$Ag for GCR-corrected data represent the propagated 2 S.D. uncertainties of the GCR correction (Table 2). Isochrons are determined using GCR-corrected data and ISOPLOT[55]. *denotes GCR-corrected data from source[11], shown for comparison but not included in the regressions.

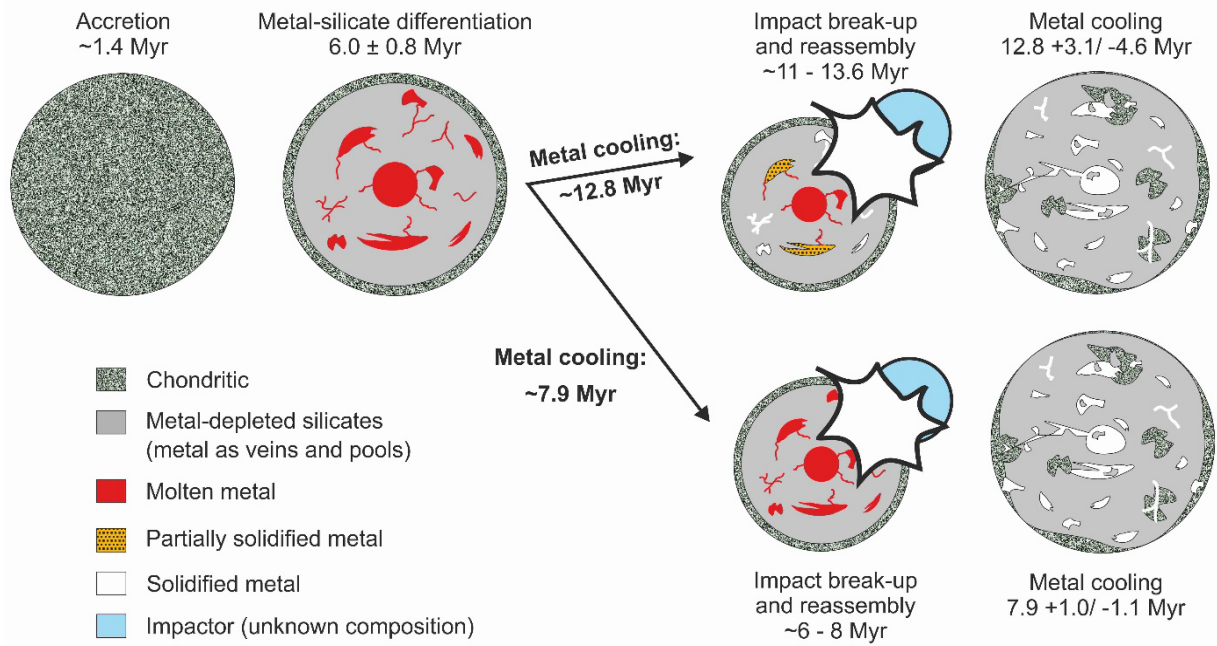

**Extended Data Figure 3.** IAB parent body evolution for both metal cooling times. In both scenarios the parent body accreted relatively late at ~1.4 Myr and then underwent limited metal-silicate differentiation at ~6.0 Myr after CAI[14]. For metal cooling at ~12.8 Myr, the catastrophic impact event occurred in the timeframe ~11 – 13.6 Myr. This was followed by fast cooling and closure of the Pd-Ag system at ~12.8 $^{+3.1}/_{-4.6}$ Myr. For metal cooling at ~7.9 Myr after CAI the body was disrupted at about 6 Myr, while at or near its peak temperature. It then cooled quickly, leading to closure of the Pd-Ag system at 7.9 $^{+1.0}/_{-1.1}$ Myr.

Supplementary Information Table 1. Summary of concentration and isotope data for Pd-Ag and Pt for IAB, IIAB and IIIAB iron meteorites.

| Meteorite | Class (and sub-group, IAB[a]) | Sample Source | Sample Identifier | Weight digested for Ag isotope measurements (g) | Pd-Ag data measured at | Ag and Pt Isotope Sampling Adjacent?[b] | Pd (ppb) | Ag (ppb) | $^{108}Pd/^{109}Ag$ | $\pm 2$ S.D. | $^{107}Ag/^{109}Ag$ | $\pm 2$ S.D. | $\varepsilon^{107}Ag_{pre-GCR-exp}$[c] | $\varepsilon^{192}Pt$ | $\pm 2$ S.D. | $\varepsilon^{194}Pt$ | $\pm 2$ S.D. | $\varepsilon^{196}Pt$ | $\pm 2$ S.D. | Ir/Pt ratio | Ir/Pt ratio source |
|---|---|---|---|---|---|---|---|---|---|---|---|---|---|---|---|---|---|---|---|---|---|
| Terrestrial | | | | | | | | | | | | | 1.079976[c] | | | | | | | | |
| Campo del Cielo | IAB-MG | Smithsonian Institution | USNM 5615 | | Taken from Theis et al. (2013) | | 3616 | 177.8 | 11.3 | 15.8 | 1.081466 | 0.6 | 18.9 | 3.5 | 0.96 | 0.73 | 0.26 | 0.15 | 0.09 | 0.09 | 0.47 | Wasson and Kallemeyn (2002) |
| Canyon Diablo[d] | IAB-MG | ETH Zurich | | | Taken from Theis et al. (2013) | No | 3888 | 53.9 | 39 | 1.4 | 1.079911 | 0.6 | 1.3 | 1.1 | -0.48 | 0.73 | 0.04 | 0.15 | 0 | 0.09 | 0.40 | Wasson and Kallemeyn (2002) |
| Odessa 1 | IAB-MG | The University of Manchester | H10.88 | | Taken from Theis et al. (2013) | | 4199 | 81.1 | 28.2 | 2.2 | 1.079998 | 0.6 | 7.2 | 1.6 | 6.26 | 0.73 | 0.39 | 0.15 | 0.31 | 0.09 | 0.41 | Wasson and Kallemeyn (2002) |
| Odessa 2 | IAB-MG | Smithsonian Institution | USNM 1418 | | Taken from Theis et al. (2013) | | 4063 | 199.8 | 10.9 | 11.7 | 1.081023 | 0.6 | 23.6 | 4.0 | 5.59 | 0.73 | 0.44 | 0.15 | 0.28 | 0.09 | 0.41 | Wasson and Kallemeyn (2002) |
| Goose Lake 1 | IAB-sLL | ETH Zurich | | | Taken from Theis et al. (2013) | | 4472 | 109.9 | 22.4 | 7.4 | 1.080559 | 0.6 | 9.0 | 2.2 | 1.52 | 0.73 | 0.15 | 0.15 | 0.08 | 0.09 | 0.36 | Wasson and Kallemeyn (2002) |
| Goose Lake 2 | IAB-sLL | ETH Zurich | | | Taken from Theis et al. (2013) | | 3881 | 116.6 | 18.5 | 8.6 | 1.080689 | 0.6 | 10.3 | 2.3 | 1.52 | 0.73 | 0.15 | 0.15 | 0.08 | 0.09 | 0.36 | Wasson and Kallemeyn (2002) |
| Toluca | IAB-sLL | Smithsonian Institution | USNM 1160 | | Taken from Theis et al. (2013) | | 4546 | 64.7 | 38.3 | 1.4 | 1.079911 | 0.6 | 3.0 | 1.3 | 2.70 | 0.73 | 0.26 | 0.15 | 0.13 | 0.09 | 0.43 | Wasson and Kallemeyn (2002) |
| Caddo County[d] | IAB-UG | ETH Zurich | | | Taken from Theis et al. (2013) | No | 4113 | 37.7 | 58.9 | 0.1 | 1.079771 | 0.6 | 0.0 | 0.9 | 0.09 | 0.73 | 0.06 | 0.15 | -0.01 | 0.09 | 0.40 | Wasson and Kallemeyn (2002) |
| Coahuila 1[e] | IIAB | | BM 54242 | 15.315 | The University of Manchester | | 1829 | 307.9 | 3.3 | 53.4 | 1.08553 | 0.6 | 53.7 | 4.8 | 0.07 | 0.49 | 0.05 | 0.17 | 0.01 | 0.07 | 0.48 - 0.53 | Pernicka and Wasson (1987); Petaev and Jacobsen (2004) |
| Coahuila 2[e] | IIAB | Smithsonian Institution | USNM 2725 | 17.29 | Carnegie Institution, Washington | No | 1708 | 251.2 | 3.8 | 50.6 | 1.085223 | 0.5 | 50.8 | 3.9 | 0.07 | 0.49 | 0.05 | 0.17 | 0.01 | 0.07 | 0.48 - 0.53 | Pernicka and Wasson (1987); Petaev and Jacobsen (2004) |
| Negrillos[f] | IIAB | Smithsonian Institution | USNM 1222 | 18.83 | Carnegie Institution, Washington | No | 1567 | 260.9 | 3.4 | 52.7 | 1.08545 | 0.5 | 50.4 | 2.9 | | | | | -0.03 | 0.05 | 1.20 - 1.43 | Petaev and Jacobsen (2004); Wasson et al. (2007) |
| North Chile 1 | IIAB | Natural History Museum, London | BM 1959, 915 | 17.266 | The University of Manchester | No | 1932 | 324.4 | 3.3 | 55 | 1.085697 | 0.6 | 63.2 | 5.7 | 0.82 | 0.83 | 0.24 | 0.04 | 0.12 | 0.08 | 0.15 - 0.16 | Pernicka and Wasson(1987); Wasson et al. (2007) |
| North Chile 2 | IIAB | Smithsonian Institution | USNM 1334 | 10.52 | Carnegie Institution, Washington | No | 1841 | 321.6 | 3.2 | 53.1 | 1.085488 | 0.5 | 61.2 | 5.6 | 0.82 | 0.83 | 0.24 | 0.04 | 0.12 | 0.08 | 0.15 - 0.16 | Pernicka and Wasson (1987); Wasson et al. (2007) |
| Sikhote Alin 1 | IIAB | Natural History Museum, London | BM 1992, M38 | 20.001 | The University of Manchester | | 2451 | 433.5 | 3.2 | 59.9 | 1.086227 | 0.6 | 80.6 | 8.7 | -0.26 | 0.73 | 0.23 | 0.15 | 0.22 | 0.09 | 0.004 | Wasson et al. (2007) |
| Sikhote Alin 2 | IIAB | Natural History Museum, London | BM 1992, M38 | 17.112 | The University of Manchester | | 2649 | 310.4 | 4.8 | 36.9 | 1.083746 | 0.6 | 51.6 | 6.2 | -0.26 | 0.73 | 0.23 | 0.15 | 0.22 | 0.09 | 0.004 | Wasson et al. (2007) |
| Boxhole | IIIAB | Natural History Museum, London | BM1938, 406 | 18.17 | The University of Manchester | | 2385 | 678.7 | 1.9 | 103.5 | 1.090933 | 0.6 | 140.6 | 13.7 | 6.83 | 0.73 | 0.28 | 0.15 | 0.25 | 0.09 | 0.62 | Wasson (1999) |
| Henbury | IIIAB | Natural History Museum, London | BM 1932, 98 | 17.243 | The University of Manchester | | 2199 | 588.8 | 2 | 77.7 | 1.088147 | 0.6 | 126.8 | 11.9 | 15.44 | 0.73 | 0.82 | 0.15 | 0.38 | 0.09 | Not available | |
| Thunda | IIIAB | Natural History Museum, London | BM 2005, M269 | 16.159 | The University of Manchester | | 3092 | 407.8 | 4.1 | 83.3 | 1.088754 | 0.6 | 87.0 | 8.2 | 0.35 | 0.73 | 0.17 | 0.15 | 0.05 | 0.09 | 0.75 - 0.81 | Pernicka and Wasson (1987); Petaev and Jacobsen (2004) |

**UNCERTAINTIES**

Column J: Uncertainties (2 S.D.) on $^{108}Pd/^{109}Ag$ ratios are estimated to be ~1 % for data collected at the University of Manchester and 5 % for data collected at the Carnegie Institution of Washington (see Extended Methods).

Column K: Uncertainties (2 S.D.)for $\varepsilon^{107}Ag$ are based on repeat analyses of reference materials (see Extended Methods).

Columns Q, S, U: Platinum isotope uncertainties (2 S.D.) are based on repeat analyses of an in-house standard (see Extended Methods).

**ADDITIONAL INFORMATION**

[a] IAB sub-groups from Wasson and Kallemeyn (2002). MG, Main Group; sLL, low Au-low Ni; UG, ungrouped / closely related to sLL

[b] 'No' denotes samples where Pt isotopes were not determined on material sampled adjacent to that for Pd-Ag isotope measurements.

[c] Terrestrial $^{107}Ag/^{109}Ag$ ratio from Schönbächler et al. (2008)

[d] Platinum isotope data taken from Hunt et al. (2018)

[e] Platinum isotope data taken from Hunt et al. (2017)

[f] Platinum isotope data taken from Kruijer et al. (2014)